\documentclass[aps,prd,twocolumn,nofootinbib]{revtex4-2}
\usepackage[colorlinks]{hyperref}
\usepackage{amsmath,mathtools}
\usepackage{epsfig}
\usepackage{graphicx}
\usepackage{dcolumn}
\usepackage{amssymb,amsmath}
\usepackage{mathrsfs}
\usepackage{epsfig,bm}
\usepackage{bm}
\usepackage{booktabs}
\usepackage{siunitx}
\begin{document}

\title{Relative enhancement of  low-mass vector-boson exchange in higher waves matrix elements: parity non-conservation in hydrogen}

\author{V. A. Dzuba}\email{v.dzuba@unsw.edu.au}
\author{V. V. Flambaum}\email{v.flambaum@unsw.edu.au}
\author{G. K. Vong}\email{g.vong@unsw.edu.au}

\affiliation{School of Physics, University of New South Wales, Sydney 2052, Australia}

\begin{abstract}
Models of unification predict additional  $Z'$ boson, which contributes to parity non-conservation  (PNC) in atoms. If  $Z'$ boson is light, ratio of $Z'$ boson contribution to the Standard Model $Z$ boson contribution to atomic PNC increases with decreasing  nuclear charge $Z$ faster than $1/Z^2$. This motivated our previous study of PNC in hydrogen and deuterium proportional to the weak interaction matrix elements $<s|W|p> $.  An enormous additional relative enhancement appears in the matrix elements between higher waves, such as $<p_{1/2,3/2} | W | d_{3/2,5/2}> $,  since $p_{3/2}$ and  $d_{3/2,5/2}$  wave functions  vanish at $r \to 0$, suppressing matrix elements of the contact $Z$ boson mediated contact electron-nucleus  interaction.  Measurements of  $<p_{1/2,3/2} | W | d_{3/2,5/2}> $ will simplify disentanglement of  the $Z'$ contribution from the Standard Model background. 
\end{abstract}

\maketitle
\section{Introduction}

Atomic parity nonconservation (PNC) continues to provide one of the most accurate low-energy tests of the electroweak interaction and a sensitive means of searching for new neutral-current physics; see, for example, the reviews in Refs.~\cite{DF-PNC12,RevPNC}. The basic mechanism is well known: weak electron--nucleus interactions admix atomic states of opposite parity and thereby induce electric-dipole amplitudes that are forbidden in the absence of parity violation. Because these amplitudes can be both measured and calculated with high precision, atomic PNC has long served as a probe of the Standard Model and of possible extensions of it, including additional neutral gauge bosons~\cite{BouchiatFayet2005,DienerGodfreyTuran2012,DFS17}.

In the case of very close levels of opposite parity, matrix elements of the weak interaction between these  states  may be measured using interference between the weak interaction and  the interaction with oscillating  electric field which also produces this mixing between close levels. This type of experiment has been done  in Dy atom \cite{Dy}. Such measurement may also be performed in hydrogen and deuterium.


Among atomic systems, hydrogen is distinguished by its exceptional theoretical simplicity. In contrast to heavy many-electron atoms, it is free from many-body electronic-structure uncertainties and allows one to study parity-violating electron--proton interactions in an especially transparent form. The possibility of observing parity violation in hydrogen was pointed out long ago by Cahn and Kane~\cite{CahnKane1977}. Subsequent work developed the formalism for hydrogen and deuterium in detail, identified the nuclear-spin-independent (NSI) weak coupling constant $C_{1p}$ and the nuclear-spin-dependent (NSD) weak coupling constant $C_{2p}$ as the relevant low-energy parameters, and emphasized the important role of metastable states and level crossings in enhancing observable effects~\cite{DunfordHolt2007,DunfordHolt2011}. More recently, laser-based and quantum-control approaches have renewed interest in hydrogen PNC as a clean source of information on the proton weak charge and on weak neutral-current couplings~\cite{RasorYost2020,LiDereviankoElliott2024}.

At the same time, parity-violating electron scattering has emerged as a complementary and conceptually clean probe of semileptonic neutral-current interactions. The proton weak charge is especially sensitive to new physics because its Standard Model value is accidentally small~\cite{ErlerKurylovRamseyMusolf2003}. The final Qweak result for parity-violating elastic $ep$ scattering agrees with the Standard Model and places strong limits on additional parity-violating semileptonic interactions~\cite{AndroicEtAl2018}. In the heavy-boson regime these bounds admit a model-independent effective-field-theory interpretation in terms of contact operators, and Ref.~\cite{CarliniVanOersPittSmith2019} gives a useful summary of the present constraints. Future and ongoing parity-violating electron-scattering programs will extend this sensitivity to extra neutral gauge bosons over a broad mass range~\cite{DevRodejohannXuZhang2021,ThomasWangWilliams2022,ThomasWangWilliams2025}.

The possibility of an additional neutral boson $Z'$ is therefore of considerable current interest. When the boson is sufficiently heavy, its exchange reduces at atomic energies to an effective local four-fermion interaction. In atomic language, the corresponding effect may then be absorbed into shifts of the conventional weak charge or weak coupling constants. This short-range regime underlies most model-independent discussions of heavy-$Z'$ constraints from parity-violation experiments~\cite{DienerGodfreyTuran2012,CarliniVanOersPittSmith2019}. If, however, the boson mass is small enough that its Compton wavelength becomes comparable to or larger than the atomic length scale, the interaction is no longer local, and one must retain the full Yukawa form of the potential. In that case, the atomic matrix elements depend on the detailed spatial overlap of the electronic wave functions with the finite-range parity-violating potential rather than only on contact densities.

This finite-range regime has already been investigated in our earlier work on heavy atoms and ions~\cite{DFS17,Sr}. Those studies provided the general framework for treating parity-violating interactions generated by a vector boson of arbitrary mass and for connecting the contact and long-range limits in a unified way. We subsequently applied this approach explicitly to hydrogen in Ref.~\cite{HPV}, where we considered parity violation induced by a $Z'$ boson of arbitrary mass for the standard $s$--$p$ mixing channels, including both NSI and NSD interactions. In that paper the new contribution was analyzed in relation to the familiar Standard Model $Z$-boson background.

The motivation of the present work is different. Here we turn to parity mixing between the $3d$ and $3p$ states of hydrogen induced by exchange of a $Z'$ boson. This case is especially attractive because there is no corresponding Standard Model $Z$-boson contribution   since $p_{3/2}$ and  $d_{3/2,5/2}$  wave functions  vanish at $r \to 0$, suppressing matrix elements of the contact $Z$ boson mediated contact electron-nucleus  interaction. As a result, any observable parity-violating effect in this channel would provide a direct signature of the new interaction, without the need to separate it from an irreducible Standard Model background. This feature makes the $3d$--$3p$ manifold an unusually clean setting for a search for new neutral-current physics.

A further advantage is that  the  $3p_{3/2}$ and  $3d_{3/2}$  levels are separated by a very small energy interval.
Here  the weak mixing amplitude is divided by a small residual energy denominator, which can strongly enhance the observable parity-violating signal. The intervals between  all $3p_{1/2,3/2}$ and  $3d_{3/2,5/2}$  energy  levels  may be  manipulated by an external magnetic field including  crossing of the levels.  In this respect the mechanism is similar to the level-crossing enhancement discussed previously in hydrogen PNC~\cite{DunfordHolt2007,DunfordHolt2011}, but in the present case it is combined with the absence of a Standard Model $Z$-boson contribution in the same mixing channel.

 Heavy-atom PNC, parity-violating electron scattering, and hydrogen spectroscopy probe related but not identical combinations of couplings~\cite{ErlerKurylovRamseyMusolf2003,AndroicEtAl2018,CarliniVanOersPittSmith2019}. Hydrogen remains particularly valuable because the relevant matrix elements may be derived analytically or semi-analytically with controlled accuracy, making the dependence on the boson mass and on the underlying axial-electron/vector-proton and vector-electron/axial-proton coupling structures especially transparent. As emphasized in Ref.~\cite{DFS17}, the ratio of the low-mass $Z'$ contribution to the Standard Model $Z$-boson contribution increases rapidly with decreasing nuclear charge, faster than $1/Z^2$, which further strengthens the case for light systems.

In this paper we extend the analysis of Ref.~\cite{HPV} to the $3d$--$3p$ manifold in hydrogen. We derive the NSI and NSD parity-violating matrix elements produced by a $Z'$ boson of arbitrary mass. The absence of a Standard Model contribution of the same type makes this system a particularly transparent probe of a new parity-violating interaction.

\section{Interaction Lagrangian for \texorpdfstring{$Z$}{Z} and \texorpdfstring{$Z'$}{Z'} bosons.}

The neutral-current  Lagrangian, including $Z$ boson field and its interaction with fermions,  is  ($\hbar=c=1$)
\begin{equation}
\mathcal L_Z = -\frac14 Z_{\mu\nu}Z^{\mu\nu} + \frac12 m_Z^2  Z_\mu Z^\mu - \frac{g}{2\cos\theta_W} Z_\mu J_Z^\mu,
\end{equation}
with  the fermion current 
\begin{equation}
J_Z^\mu = \sum_f \bar f\gamma^\mu\bigl(g_f^V-g_f^A\gamma_5\bigr)f.
\end{equation}
 Here $m_Z$ is the $Z$ boson mass and $g$ is the universal interaction constant of the unified electroweak theory where   proton electric charge is 
$e=g \sin\theta_W$ and  $\theta_W$ is the Weinberg angle. Dimensionless fermion interaction constants $g_f^V$ and $g_f^A$ for proton and electron are presented below;  $\gamma^\mu$ and $\gamma_5$ are the Dirac matrices.

\subsection{Nuclear-spin-independent PV interaction}

The parity-violating nuclear-spin-independent part of the interaction  comes from the electron axial current and the proton vector current.  For the SM $Z$-boson (without radiative corrections) 
\begin{equation}
 g_e^A=-\frac12,
 \qquad
 g_p^V=\frac12-2\sin^2\theta_W,
 \end{equation}
Proton weak charge, which determines strength of PNC electron - proton interaction, is equal to 
\begin{equation}
 Q_W^p=-4g_e^A g_p^V = 1-4\sin^2\theta_W.
\end{equation}
Inclusion of radiative corrections gives $ Q_W^p = 0.0705$ 
\cite{SM}.  Corrections produced by the dispersive parity violating interaction may also be included into the definition of the weak charge giving $ Q_W^p = 0.0719$ \cite{Samsonov}. The Qweak experimental result is
$Q_W^{p,\mathrm{exp}}=0.0719(45)$~\cite{AndroicEtAl2018}.

Then we may write the NSI part of the  $Z$-exchange parity-violating Hamiltonian as
\begin{equation}
V_Z(r) = -\frac{G_F}{2\sqrt2\,}\,Q_W^p\,m_Z^2 \Phi(m_Z,r)\,\gamma_5,
\label{eq:Vzfull}
\end{equation}
where $G_F$
  is the Fermi constant of the weak interaction,
  \begin{equation}
\Phi(m,r) =  \frac{1}{4\pi} \frac{e^{-m  r}}{r}.
\end{equation}
In the case of the large mediator mass, $m \to \infty$ , 
\begin{equation}
\Phi(m,r) =  \frac{1}{m^2} \delta^3(r)\,.
\end{equation}

For a generic parity-violating $Z'$ boson we absorb the model-dependent gauge coupling and charges into two effective couplings $g_{eA}'$ and $g_{pV}'$ and present  NSI interaction potential as 
\begin{equation}
V_{Z'}(r) = g_{eA}' g_{pV}'\,\Phi(m_{Z'},r)\,\gamma_5 \,.
\label{eq:Vzprimepot}
\end{equation}

\subsection{ \texorpdfstring{$Z'$}{Z'} boson matrix elements}

We write the Dirac orbital as
\begin{equation}
\psi_{n\kappa m}(\mathbf r)
= \frac1r
\begin{pmatrix}
 g_{n\kappa}(r)\,\Omega_{\kappa m}(\hat{\mathbf r}) \\
 i \alpha f_{n\kappa}(r)\,\Omega_{-\kappa m}(\hat{\mathbf r})
\end{pmatrix}.
\end{equation}
Here $\alpha$ is the fine structure constant. For the \(3p_{3/2}\) and \(3d_{3/2}\) states one has \(\kappa_{p_{3/2}}=-2\) and
\(\kappa_{d_{3/2}}=+2\).  In the nonrelativistic limit
\begin{align}
g_{3p}(r') &= r'R_{31}(r'), &\\
g_{3d}(r') &= r'R_{32}(r'),\\
f_{3p}(r') &= \frac1{2}\left(\frac{d}{dr'}-\frac{2}{r'}\right)g_{3p}(r'), &\\
f_{3d}(r') &= \frac1{2}\left(\frac{d}{dr'}+\frac{2}{r'}\right)g_{3d}(r'),
\end{align}
with
\begin{align}
R_{31}(r') &= \frac{4}{81\sqrt6}\,r'(6-r')e^{-r'/3},\\
R_{32}(r') &= \frac{4}{81\sqrt{30}}\,r'^2 e^{-r'/3}.
\end{align}
Here $r'= r/a_B$, where $a_B=1/(m_e \alpha)$ is the Bohr radius.

\subsection{NSI \(3p_{3/2}\)--\(3d_{3/2}\) matrix element}

Since \(\kappa_f=-\kappa_i\), the NSI \(\gamma_5\) matrix element has the same radial structure
as in the \(ns_{1/2}\)--\(np_{1/2}\) case:
\begin{eqnarray}
\nonumber
\mathcal M_{pd}^{Z',\rm NSI}
\equiv
\langle 3p_{3/2}m|V_{Z'}^{\rm NSI}|3d_{3/2}m\rangle
=\\
i g'_{eA}g'_{pV}\alpha \int_0^\infty dr\,
\bigl(g_{3p}f_{3d}-f_{3p}g_{3d}\bigr)\Phi(m_{Z'},r).
\end{eqnarray}
Evaluation of the radial integral gives
\begin{equation}
\mathcal M_{pd}^{Z',\rm NSI}
=
i\,\frac{4\sqrt5}{135\pi a_B}\,
g'_{eA}g'_{pV}\,\alpha 
\frac{15\mu+2}{(3\mu+2)^5}
\end{equation}
for any \(m=\pm 1/2,\pm 3/2\).  Here  $\mu=m_{Z'} a_B= {m_Z'}/( \alpha m_e)$.


For a heavy SM \(Z\) boson the contact matrix element between \(3p\) and \(3d\) vanishes for a
point proton.  It is therefore more useful to compare \(\mathcal M_{pd}^{Z',\rm NSI}\) with the standard model 
\(2s\)--\(2p_{1/2}\)  matrix element,
\begin{equation}
\mathcal M_{2}^{Z,\rm NSI}
=
i\alpha \frac{G_F Q_W^p\sqrt3}{64\pi\sqrt2a_B^3}.
\end{equation}
The ratio is
\begin{equation}
\frac{\mathcal M_{pd}^{Z',\rm NSI}}{\mathcal M_2^{Z,\rm NSI}}
=
g'_{eA}g'_{pV}\,
\frac{256\sqrt{30 }a_B^2}{405\,G_FQ_W^p}\,
\frac{15\mu+2}{(3\mu+2)^5}
\end{equation}
with limiting forms
\begin{align}
\mu\to0:\quad
\frac{\mathcal M_{pd}^{Z',\rm NSI}}{\mathcal M_2^{Z,\rm NSI}}
&\to
g'_{eA}g'_{pV}\,
\frac{16\sqrt{30}a_B^2}{405\,G_FQ_W^p},
\\
\mu\to\infty:\quad
\frac{\mathcal M_{pd}^{Z',\rm NSI}}{\mathcal M_2^{Z,\rm NSI}}
&\to
g'_{eA}g'_{pV}\,
\frac{256\sqrt{30}a_B^2}{6561
\,G_FQ_W^p}\,\mu^{-4}.
\end{align}

Numerically, for \(Q_W^p=0.0719\),
\begin{equation}
\left|
\frac{\mathcal M_{pd}^{Z',\rm NSI}}{\mathcal M_2^{Z,\rm NSI}}
\right|_{\mu\to 0}
=
1.856\times 10^{16}\,|g'_{eA}g'_{pV}|.
\end{equation}

\subsection{Nuclear-spin-dependent interaction}

With the current convention
\(\bar f\gamma^\mu(g_f^V-g_f^A\gamma_5)f\), the vector-electron--axial-proton
part of the finite-range interaction is written as
\begin{equation}
V_{Z'}^{\rm NSD}(r)
=
-\frac{g'_{eV}g'_{pA}}{I}\,
{\bm\alpha}\cdot{\bm I}\,\Phi(m_{Z'},r).
\label{eq:VzprimeNSD_checked}
\end{equation}
The overall sign is convention dependent and may equivalently be absorbed into
the definition of \(g'_{pA}\); it is fixed here so that all NSD amplitudes below
are mutually consistent.  For comparison, the Standard-Model contact
Hamiltonian is
\begin{equation}
H_Z^{\rm NSD}
=
\frac{G_F}{\sqrt2}\frac{\varkappa_p}{I}\,
{\bm\alpha}\cdot{\bm I}\,\delta^3(\mathbf r),
\label{eq:HSDcontact_checked}
\end{equation}
where we use \(\varkappa_p=0.043\), the radiatively corrected
electron-vector--proton-axial coupling conventionally quoted for hydrogen
when the proton anapole contribution is omitted~\cite{DunfordHolt2007}.

\subsection{NSD \(3p_{3/2}\)--\(3d_{3/2}\) matrix element}

For hydrogen in the stretched hyperfine state \(I=1/2\), \(m_I=1/2\), \(m_j=3/2\),
one has \(({\bm\alpha}\cdot{\bm I})/I\to \alpha_z\).  The relevant angular factors are
\begin{equation}
\langle \Omega_{-2,3/2}|\sigma_z|\Omega_{-2,3/2}\rangle = 1,
\qquad
\langle \Omega_{2,3/2}|\sigma_z|\Omega_{2,3/2}\rangle = -\frac35.
\end{equation}
Hence
\begin{eqnarray}
\nonumber
\mathcal M_{pd,z}^{Z',\rm NSD}
\equiv
\langle 3p_{3/2},m_j=\tfrac32|V_{Z'}^{\rm NSD}|3d_{3/2},m_j=\tfrac32\rangle
=\\
-i\,g'_{eV}g'_{pA}\alpha
\int_0^\infty dr\,
\left(g_{3p}f_{3d}+\frac35 f_{3p}g_{3d}\right)\Phi(m_{Z'},r),\,\,\,\,\,
\end{eqnarray}
which yields
\begin{equation}
\mathcal M_{pd,z}^{Z',\rm NSD}
=
-i\,\frac{4\sqrt5}{675\pi a_B}\,
g'_{eV}g'_{pA}\alpha
\frac{225\mu^2+36\mu+4}{(3\mu+2)^6}
\end{equation}
for the \(z\) component in the stretched state.


Comparing with the standard \(2s\)--\(2p_{1/2}\) NSD matrix element,
\begin{equation}
\mathcal M_2^{Z,\rm NSD}
=
-i\alpha \frac{\sqrt3 \,G_F\varkappa_p}{32\pi\sqrt2\,a_B^3},
\end{equation}
we obtain
\begin{equation}
\frac{\mathcal M_{pd,z}^{Z',\rm NSD}}{\mathcal M_2^{Z,\rm NSD}}
=
g'_{eV}g'_{pA}\,
\frac{128\sqrt{30}\,a_B^2}{2025\,G_F\varkappa_p}\,
\frac{225\mu^2+36\mu+4}{(3\mu+2)^6}
\end{equation}
with the low-mass limit
\begin{equation}
\left|
\frac{\mathcal M_{pd,z}^{Z',\rm NSD}}{\mathcal M_2^{Z,\rm NSD}}
\right|_{\mu\to 0}
=
3.103\times 10^{15}\,|g'_{eV}g'_{pA}|.
\end{equation}

The stretched states corresponds to the hyperfine component F=2. The $F=1$ and $F=2$ hyperfine amplitudes are related by (see Appendix)
\begin{eqnarray}
\nonumber
\langle 3p_{3/2},F=1,F_z|V_{Z'}^{\rm NSD}|3d_{3/2},F=1,F_z\rangle
=\\
-\frac53\,
\langle 3p_{3/2},F=2,F_z|V_{Z'}^{\rm NSD}|3d_{3/2},F=2,F_z\rangle.
\end{eqnarray}

We will also need matrix elements of NSD operator between $3p_{3/2}$ and $3d_{5/2}$ states with F=2
\begin{eqnarray}
\nonumber
\langle 3p_{3/2},F=2,F_z|V_{Z'}^{\rm NSD}|3d_{5/2},F=2,F_z\rangle
=\\
-i\,\alpha\,\frac{32\sqrt{30}}{675\pi a_B}\,g_{eV}'g_{pA}'\,
\frac{9\mu+1}{(3\mu+2)^6}
\label{eq:F_p32d52_main}
\end{eqnarray}
and between $3p_{1/2}$ and $3d_{3/2}$ states with F=1
\begin{eqnarray}
\nonumber
\langle 3p_{1/2},F=1,F_z|V_{Z'}^{\rm NSD}|3d_{3/2},F=1,F_z\rangle
=\\
-i\,\alpha\,\frac{16\sqrt{10}}{405\pi a_B}\,g_{eV}'g_{pA}'\,
\frac{45\mu^2+18\mu+2}{(3\mu+2)^6}
\label{eq:F_p12d32_main}
\end{eqnarray}

 \subsection{Parity-violating E1 amplitudes and magnetic-field effect}
 \label{sec:pnc_amplitudes}

For a \(3d_{j}\) final state mixed with \(3p_{3/2}\),  $j=3/2,5/2$,  the induced E1 amplitude is
\begin{equation}
E_{\rm PNC}(E)
=
\frac{\langle 3d_{j}|V_{Z'}| 3p_{3/2} \rangle\, \langle 3p_{3/2}|D_q|n_s s_{1/2}\rangle
}{E- E_{3p_{3/2}} + i  \Gamma_{3p_{3/2}}/2},
\end{equation}
where the excitation energy $E \approx  E_{3d_{j}}$\footnote{Energy and width of $3d_j$ state  appear explicitly  if we multiply  this amplitude by a factor describing decay of $3d_j$ state to a final state $f$,  $\langle f |D_q|3d_{j}\rangle/( E- E_{3d_{j}} + i \Gamma_{3d_{j}}/2)$.}. The E1  matrix elements entering this and the following expressions are given in Appendix~\ref{app:E1E2}. The energy interval  $E_{3d_{j}} -  E_{3p_{3/2}}$  can be tuned by a magnetic field.


The experimental zero-field interval
\[
\Delta_0(3p_{3/2}-3d_{3/2}) = 5.5(9)\ {\rm MHz},
\]
is smaller than the widths of the states
\begin{align}
\Gamma_{3p_{3/2}}/h &= 30.21\ {\rm MHz},
\\
\Gamma_{3d_{3/2}}/h &= 10.30\ {\rm MHz}.
\end{align}
The resonances $3p_{3/2}$ and  $3d_{3/2}$ overlap and this makes experiment to measure $E_{\rm PNC}$ in the transitions
$1s,2s$ -   $3d_{3/2}$ practically impossible. A more realistic possibility may  be to measure interference between electric and weak amplitudes between the states  $3p_{3/2}$ and  $3d_{3/2}$ as it was done in the Dy experiment \cite{Dy}.   The resonances $3p_{3/2}$ and  $3d_{5/2}$ do not overlap and here the  measurement of NSD  $E_{\rm PNC}$ in the transitions
$1s,2s$ -   $3d_{5/2}$ is not excluded. 

In the linear Zeeman approximation with Land\'e factors
\[
g(3p_{3/2})=\frac43,\qquad g(3d_{3/2})=\frac45,
\]
the crossing field for the stretched \(m_j=3/2\) components of  $3p_{3/2}$ and  $3d_{3/2}$ states is about
\begin{equation}
B_c(3p_{3/2},3d_{3/2})
=
\frac{\Delta_0}{\mu_B(g_{3p_{3/2}}-g_{3d_{3/2}})\,(3/2)}
=
4.9\ {\rm G}
\end{equation}
using \(\mu_B/h=13.99624555\ {\rm GHz/T}\). Since
the hyperfine splittings are comparable to this very small electronic
interval, a prediction for a specified hyperfine component must include the
hyperfine Hamiltonian.

The energy intervals between the levels  $3p_{3/2}$ and  $3d_{5/2}$ , and between $3p_{1/2}$ and  $3d_{3/2}$, may be reduced by magnetic field which  could enhance their mixing and PNC amplitude. However, the magnetic interaction in this case is comparable to the fine structure interval and this leads to a nonlinear dependence  of the energy intervals and wave functions on the magnetic field.    The calculations indicate that no significant enhancement may be achieved.


The \(2s\)--\(2p_{1/2}\) interval is 
\[
\Delta_0(2s_{1/2}-2p_{1/2}) = 1057.845(3)\ {\rm MHz},
\]
and its crossing occurs near \(0.112\ {\rm T}\).  

Using the natural width of the $2p_{1/2}$ level,
\[
\Gamma_{2p_{1/2}}/h \approx 99.7\ {\rm MHz},
\]
the enhancement factor  
\[
K_{\max}(2s-2p_{1/2})\lesssim \frac{\Delta_0}{\Gamma_{2p_{1/2}}}=10
\]
The different magnetic-field dependence of the \(2s\)--\(2p\) and
\(3d\)--\(3p\) contributions to $E_{PNC}$  may be used to separate them in
\(2s\)--\(3d\) measurements by varying magnetic field, provided that the full field-dressed
Hamiltonian is used.


\section{Deuterium} 

Deuterium PNC experiments are of independent interest since they may give us interaction constant of $Z'$ boson with neutron. Energy intervals, widths and electromagnetic amplitudes in deuterium are very close to that in hydrogen.  Therefore,  to find results for $Z'$ in the NSI case it is sufficient to replace  the proton interaction constant $g'_{pV}$  by the deuterium interaction constant $g'_{dV} = g'_{pV}+g'_{nV}$. 

 The NSD rescaling is less direct.  The deuteron has spin
\(I=1\). In a zero approximation spins of proton and neutron in deuterium are parallel. However,  proton and neutron spin expectation values are modified by the
deuteron \(D\)-state admixture. Thus
\(g'_{dA}\simeq g'_{pA}+g'_{nA}\) is only a leading \(S\)-wave estimate.
Also,  the hyperfine recoupling coefficients are
different from those for hydrogen.


Within the standard model,   deuterium weak charge $Q^d_W =-0.98207 N + 0.071918 Z=0.910$  is dominated by the neutron contribution and significantly exceeds the  proton weak charge.
The standard model value  of $\varkappa_d$ is very small since $\varkappa_n \approx - \varkappa_p$.  A significant contribution is given by the nuclear anapole moment  \cite{anapole}. However, in the point-like nucleus approximation this contribution, as well as contribution of $Z$ boson, vanishes in $p-d$ matrix elements.


The angular reduction coefficient for deuterium with $I=1$ are different from  that for hydrogen with $I=1/2$.  Here we give only the resulting matrix elements.  All amplitudes below are diagonal in $F$ and $F_z$ and are
independent of $F_z$.

For the NSI $2s_{1/2}$--$2p_{1/2}$ matrix element, the allowed hyperfine
values are $F=1/2,3/2$, and
\begin{multline}
\mathcal M_{2,d}^{Z',\rm NSI}
=
 i\,\frac{\sqrt3\,\alpha}{96\pi a_B}\,
 g'_{eA}g'_{dV}\,
 \frac{3\mu+1}{(\mu+1)^3},
\label{eq:D_NSI_2s2p_Zp}
\end{multline}
The corresponding Standard-Model contact matrix element is
\begin{equation}
\mathcal M_{2,d}^{Z,\rm NSI}
=
 i\,\alpha\,
 \frac{\sqrt3\,G_FQ_W^d}{64\pi\sqrt2\,a_B^3},
\label{eq:D_NSI_2s2p_Z}
\end{equation}

For the NSI $3p_{3/2}$--$3d_{3/2}$ matrix element,
$F=1/2,3/2,5/2$, and
\begin{multline}
\mathcal M_{pd,d}^{Z',\rm NSI}
=
 i\,\frac{4\sqrt5\,\alpha}{135\pi a_B}\,
 g'_{eA}g'_{dV}\,
 \frac{15\mu+2}{(3\mu+2)^5},
\label{eq:D_NSI_p32d32}
\end{multline}

For compactness, define
\begin{equation}
\mathcal A_d
\equiv
\frac{\alpha}{\pi a_B}\,g'_{eV}g'_{dA},
\label{eq:D_Ad}
\end{equation}
and
\begin{align}
\nonumber
X_{33}(\mu)
&=\frac{225\mu^2+36\mu+4}{(3\mu+2)^6},
&
X_{35}(\mu)
&=\frac{9\mu+1}{(3\mu+2)^6},
\\
X_{13}(\mu)
&=\frac{45\mu^2+18\mu+2}{(3\mu+2)^6}.
\label{eq:D_radial_functions}
\end{align}

For the NSD $3p_{3/2}$--$3d_{3/2}$ matrix elements the results for $F=1/2,3/2,5/2$ are
\begin{equation}
\begin{aligned}
\mathcal M_{33,d}^{(F=1/2)}
&=+i\,\frac{4\sqrt5}{405}\,\mathcal A_d X_{33},
\\
\mathcal M_{33,d}^{(F=3/2)}
&=+i\,\frac{8\sqrt5}{2025}\,\mathcal A_d X_{33},
\\
\mathcal M_{33,d}^{(F=5/2)}
&=-i\,\frac{4\sqrt5}{675}\,\mathcal A_d X_{33}.
\end{aligned}
\label{eq:D_NSD_p32d32_allF}
\end{equation}

For the NSD $3p_{3/2}$--$3d_{5/2}$ matrix elements 
the common hyperfine values are $F=3/2,5/2$, and
\begin{equation}
\begin{aligned}
\mathcal M_{35,d}^{(F=3/2)}
&=-i\,\frac{16\sqrt5}{225}\,\mathcal A_d X_{35},
\\
\mathcal M_{35,d}^{(F=5/2)}
&=-i\,\frac{16\sqrt{70}}{675}\,\mathcal A_d X_{35}.
\end{aligned}
\label{eq:D_NSD_p32d52_allF}
\end{equation}

Finally, for the NSD $3p_{1/2}$--$3d_{3/2}$ matrix elements 
for $F=1/2,3/2$,
\begin{equation}
\begin{aligned}
\mathcal M_{13,d}^{(F=1/2)}
&=-i\,\frac{8\sqrt{10}}{405}\,\mathcal A_d X_{13},
\\
\mathcal M_{13,d}^{(F=3/2)}
&=-i\,\frac{8}{81}\,\mathcal A_d X_{13}.
\end{aligned}
\label{eq:D_NSD_p12d32_allF}
\end{equation}

\begin{acknowledgments}

This work was supported by the Australian Research Council Grants No. DP230101058.
\end{acknowledgments}

\appendix 
\section{Appendix} 
\subsection{$3p_{3/2}$--$3d_{3/2}$ NSD matrix elements in the hyperfine basis}

For the nuclear-spin-dependent interaction it is convenient to pass from the
uncoupled basis to the hyperfine basis
\begin{equation}
|n\kappa;FF_z\rangle
=
\sum_{m_j,m_I}
C^{FF_z}_{jm_j,Im_I}
|n\kappa jm_j\rangle |Im_I\rangle,
\qquad I=\frac12,
\end{equation}
where ${\bf F}={\bf j}+{\bf I}$.  Since
\begin{equation}
V_{Z'}^{\rm NSD}(r)=\frac{g_{eV}'g_{pA}'}{I}\,{\bm \alpha}\cdot{\bm I}\,\Phi(m_{Z'},r)
\end{equation}
is a scalar in the total angular-momentum space, its matrix elements are diagonal in
$F$ and $F_z$ and do not depend on $F_z$.

Introducing the electronic rank-1 operator
\begin{equation}
T^{(1)}_q = g_{eV}'g_{pA}'\,\alpha_q\,\Phi(m_{Z'},r),
\end{equation}
one may write
\begin{equation}
V_{Z'}^{\rm NSD}=\frac1I\sum_q (-1)^q T^{(1)}_q I_{-q}.
\end{equation}
The hyperfine matrix element is then obtained from the electronic reduced matrix element as
\begin{equation}
\mathcal M^{(F)}_{ab}
=
(-1)^{F+j_b+I}
\begin{Bmatrix}
 j_a & j_b & 1 \\
 I   & I   & F
\end{Bmatrix}
\frac{\langle I\|I\|I\rangle}{I}
\langle a j_a\|T^{(1)}\|b j_b\rangle,
\label{eq:HFrecouple_p32d32}
\end{equation}
with
\begin{equation}
\frac{\langle I\|I\|I\rangle}{I}=\sqrt6
\qquad (I=\tfrac12).
\end{equation}

For the $3p_{3/2}$--$3d_{3/2}$ channel the electronic reduced matrix element derived above is
\begin{eqnarray} \label{eq:redT_p32d32_hf}
&&\langle 3p_{3/2}\|T^{(1)}\|3d_{3/2}\rangle = \\
&&-i\,\frac{8\sqrt3}{405\pi a_B}\, g'_{eV}g'_{pA}\alpha \, \frac{225\mu^2+36\mu+4}{(3\mu+2)^6}. \nonumber
\end{eqnarray}

For $F=2$ one finds
\begin{equation}
(-1)^{F+j_b+I}
\begin{Bmatrix}
\frac32 & \frac32 & 1 \\
\frac12 & \frac12 & 2
\end{Bmatrix}
\sqrt6
=
\frac{\sqrt{15}}{10}.
\end{equation}
Therefore, for any $F_z=-2,-1,0,1,2$,
\begin{multline}
\langle 3p_{3/2},F{=}2,F_z|V_{Z'}^{\rm NSD}|3d_{3/2},F{=}2,F_z\rangle
\\
= -i \alpha \,\frac{4\sqrt5}{675\pi a_B}\,
g'_{eV}g'_{pA}\,
\frac{225\mu^2+36\mu+4}{(3\mu+2)^6}
\label{eq:F2_p32d32}
\end{multline}

which coincides with the stretched-state result, as expected.

For $F=1$ the recoupling coefficient is
\begin{equation}
(-1)^{F+j_b+I}
\begin{Bmatrix}
\frac32 & \frac32 & 1 \\
\frac12 & \frac12 & 1
\end{Bmatrix}
\sqrt6
=
-\frac{\sqrt{15}}{6}.
\end{equation}
Hence, for any $F_z=-1,0,1$,

\begin{multline}
\langle 3p_{3/2},F{=}1,F_z|V_{Z'}^{\rm NSD}|3d_{3/2},F{=}1,F_z\rangle
\\
= +i \alpha \,\frac{4\sqrt5}{405\pi a_B}\,
g'_{eV}g'_{pA}\,
\frac{225\mu^2+36\mu+4}{(3\mu+2)^6}
\label{eq:F1_p32d32}
\end{multline}

which is also independent of $F_z$.

Thus the $F=1$ and $F=2$ hyperfine amplitudes are related by
\begin{multline}
   \langle 3p_{3/2},F=1,F_z|V_{Z'}^{\rm NSD}|3d_{3/2},F=1,F_z\rangle
   \\
=
-\frac53\,
\langle 3p_{3/2},F=2,F_z|V_{Z'}^{\rm NSD}|3d_{3/2},F=2,F_z\rangle.
\end{multline}

As rank-0 reduced matrix elements in the hyperfine space one has
\begin{multline}
\langle 3p_{3/2},F{=}2\|V_{Z'}^{\rm NSD}\|3d_{3/2},F{=}2\rangle
\\
= -i \alpha \,\frac{4}{135\pi a_B}\,
g'_{eV}g'_{pA}\,
\frac{225\mu^2+36\mu+4}{(3\mu+2)^6},
\label{eq:F2_red}
\end{multline}
\begin{multline}
\langle 3p_{3/2},F{=}1\|V_{Z'}^{\rm NSD}\|3d_{3/2},F{=}1\rangle
\\
= +i \alpha \,\frac{4\sqrt{15}}{405\pi a_B}\,
g'_{eV}g'_{pA}\,
\frac{225\mu^2+36\mu+4}{(3\mu+2)^6}.
\label{eq:F1_red}
\end{multline}

We have checked explicitly by direct summation over the uncoupled basis
$|jm_j\rangle|Im_I\rangle$ that the matrix elements in
Eqs.~\eqref{eq:F2_p32d32} and \eqref{eq:F1_p32d32} are independent of $F_z$,
as required for a scalar operator in the hyperfine basis.


For the $3p_{3/2}$ and $3d_{5/2}$ states one has $\kappa_{p_{3/2}}=-2$ and
$\kappa_{d_{5/2}}=-3$. In the nonrelativistic limit the upper radial functions are
again given by
\begin{align}
 g_{3p}(r')&=r'R_{31}(r'), & g_{3d}(r')&=r'R_{32}(r'),
\end{align}
while the small component of the $3p_{3/2}$ state is
\begin{equation}
 f_{3p_{3/2}}(r')=\frac{1}{2}\left(\frac{d}{dr'}-\frac{2}{r'}\right)g_{3p}(r').
\end{equation}
In this channel only the lower-upper term contributes, since
$l(-\kappa_{p_{3/2}})=2=l(d_{5/2})$, whereas
$l(p_{3/2})=1\neq l(-\kappa_{d_{5/2}})=3$.
For example, for $m_j=3/2$ one finds
\begin{equation}
\langle \Omega_{2,3/2}|\sigma_0|\Omega_{-3,3/2}\rangle=-\frac45.
\end{equation}
The relevant radial integral is
\begin{eqnarray} \label{eq:radint}
&&\int_0^\infty dr\,f_{3p_{3/2}}(r)\,g_{3d}(r)\,\Phi(m_{Z'},r) = \\
&&- \frac{8\sqrt5\,(9\mu+1)}{135\pi a_B\,(3\mu+2)^6}. \nonumber
\end{eqnarray}
Using the Wigner-Eckart theorem for the electronic operator,
\begin{equation}
\langle 3p_{3/2}\|T^{(1)}\|3d_{5/2}\rangle
=
-i \alpha \,\frac{32\sqrt3}{135\pi a_B}\,g_{eV}'g_{pA}'\,
\frac{9\mu+1}{(3\mu+2)^6}
\label{eq:redT_p32d52}
\end{equation}
and Eq.~\eqref{eq:HFrecouple_p32d32} gives
\begin{equation}
(-1)^{F+j_b+I}
\begin{Bmatrix}
\frac32 & \frac52 & 1 \\
\frac12 & \frac12 & 2
\end{Bmatrix}
\sqrt6
=\frac{\sqrt{10}}{5}.
\end{equation}
Therefore, for any $F_z=-2,-1,0,1,2$,
\begin{multline}
\langle 3p_{3/2},F=2,F_z|V_{Z'}^{\rm NSD}|3d_{5/2},F=2,F_z\rangle
\\
=-i\,\alpha\,\frac{32\sqrt{30}}{675\pi a_B}\,g_{eV}'g_{pA}'\,
\frac{9\mu+1}{(3\mu+2)^6}
\label{eq:F_p32d52}
\end{multline}
which is manifestly independent of $F_z$.
As a rank-0 reduced matrix element in the hyperfine space,
\begin{multline}
\langle 3p_{3/2},F=2\|V_{Z'}^{\rm NSD}\|3d_{5/2},F=2\rangle
\\
=-i \alpha\,\frac{32\sqrt6}{135\pi a_B}\,g_{eV}'g_{pA}'\,
\frac{9\mu+1}{(3\mu+2)^6}.
\end{multline}

\subsubsection{$3p_{1/2}$--$3d_{3/2}$, $F=1$}

For the $3p_{1/2}$ and $3d_{3/2}$ states one has $\kappa_{p_{1/2}}=+1$ and
$\kappa_{d_{3/2}}=+2$. In this case only the upper-lower term contributes, since
$l(p_{1/2})=1=l(-\kappa_{d_{3/2}})$, whereas
$l(-\kappa_{p_{1/2}})=0\neq l(d_{3/2})=2$.
The small component of the $3d_{3/2}$ state is
\begin{equation}
 f_{3d_{3/2}}(r')=\frac{1}{2}\left(\frac{d}{dr'}+\frac{2}{r'}\right)g_{3d}(r').
\end{equation}
For $m_j=1/2$ one finds
\begin{equation}
\langle \Omega_{1,1/2}|\sigma_0|\Omega_{-2,1/2}\rangle=-\frac{2\sqrt2}{3}.
\end{equation}
The radial integral is
\begin{eqnarray}
\int_0^\infty dr\,g_{3p}(r)f_{3d_{3/2}}(r)\Phi(m_{Z'},r) = \\
 \frac{4\sqrt5}{135\pi a_B} \frac{45\mu^2+18\mu+2}{(3\mu+2)^6}. \nonumber
\end{eqnarray}
Hence the electronic reduced matrix element is
\begin{eqnarray} \label{eq:redT_p12d32}
&&\langle 3p_{1/2}\|T^{(1)}\|3d_{3/2}\rangle = \\
&&-i \alpha \,\frac{16\sqrt{15}}{405\pi a_B}\,g_{eV}'g_{pA}'\,
\frac{45\mu^2+18\mu+2}{(3\mu+2)^6} \nonumber
\end{eqnarray}
and for the hyperfine matrix element one has
\begin{equation}
(-1)^{F+j_b+I}
\begin{Bmatrix}
\frac12 & \frac32 & 1 \\
\frac12 & \frac12 & 1
\end{Bmatrix}
\sqrt6
=\frac{\sqrt6}{3}.
\end{equation}
Therefore, for any $F_z=-1,0,1$,

\begin{multline}
\langle 3p_{1/2},F{=}1,F_z|V_{Z'}^{\rm NSD}|3d_{3/2},F{=}1,F_z\rangle
\\
=-i\alpha \,\frac{16\sqrt{10}}{405\pi a_B}\,g_{eV}'g_{pA}'\,
\frac{45\mu^2+18\mu+2}{(3\mu+2)^6}\,
\label{eq:F_p12d32}
\end{multline}
which again does not depend on $F_z$.
The corresponding reduced matrix element in the hyperfine space is
\begin{multline}
\langle 3p_{1/2},F{=}1\|V_{Z'}^{\rm NSD}\|3d_{3/2},F{=}1\rangle
\\
=-i \alpha \,\frac{16\sqrt{30}}{405\pi a_B}\,g_{eV}'g_{pA}'\,
\frac{45\mu^2+18\mu+2}{(3\mu+2)^6}.
\end{multline}

We have checked explicitly by direct summation over the uncoupled states
$|jm_j\rangle|Im_I\rangle$ that Eqs.~\eqref{eq:F_p32d52} and \eqref{eq:F_p12d32}
are independent of $F_z$, as required for a scalar operator in the hyperfine basis.

\subsection{Electric-dipole and electric-quadrupole matrix elements}
\label{app:E1E2}

The PNC amplitudes $E1_{\rm PNC}$ in Sec.~\ref{sec:pnc_amplitudes} are proportional to the weak mixing coefficient between opposite parity states times ordinary E1 amplitudes. One also needs E1 amplitudes to find mixing of opposite parity states by an electric field and background E2 amplitudes between $s$ and $d$ states. The E1 and E2 amplitudes presented below are standard hydrogenic results~\cite{BetheSalpeter1957}, which we collect here for completeness.

The standard hydrogenic radial integrals are quoted below in atomic units, i.e.\
 the E1 matrix elements are given in units of $a_B$ and the E2 matrix elements in units of $a_B^2$.\
 The numerical coefficients are therefore unchanged from the atomic-unit formulas. The standard hydrogenic radial integrals are
\begin{align}
\int_0^\infty R_{10}(r)\,r\,R_{31}(r)\,r^2dr &= \frac{27\sqrt6}{128},
\\
\int_0^\infty R_{20}(r)\,r\,R_{31}(r)\,r^2dr &= \frac{27648\sqrt3}{15625},
\\
\int_0^\infty R_{10}(r)\,r^2\,R_{32}(r)\,r^2dr &= \frac{81\sqrt{30}}{256},
\\
\int_0^\infty R_{20}(r)\,r^2\,R_{32}(r)\,r^2dr &= -\frac{5308416\sqrt{15}}{390625}.
\end{align}
From these one finds the reduced E1 matrix elements
\begin{align}
\langle 3p_{3/2}\|rC^{(1)}\|1s_{1/2}\rangle
= \frac{27\sqrt2}{64}
\\
\langle 3p_{3/2}\|rC^{(1)}\|2s_{1/2}\rangle
= \frac{55296}{15625}
\end{align}
For the \(q=0\) (\(z\)-polarized) E1 components one has
\begin{align}
\langle 3p_{3/2},\tfrac12|z|1s_{1/2},\tfrac12\rangle &= \frac{9\sqrt3}{64},
\\
\langle 3p_{3/2},\tfrac12|z|2s_{1/2},\tfrac12\rangle &= \frac{9216\sqrt6}{15625}.
\end{align}
For the stretched \(m_j=3/2\) final state, however, the relevant E1 operator is \(q=+1\),
not \(q=0\), because an \(s_{1/2}\) initial state cannot reach \(m_j=3/2\) with \(\Delta m=0\).  The
needed circular components are
\begin{align}
\langle 3p_{3/2},\tfrac32|r_{+1}|1s_{1/2},\tfrac12\rangle &= \frac{27\sqrt2}{128},
\\
\langle 3p_{3/2},\tfrac32|r_{+1}|2s_{1/2},\tfrac12\rangle &= \frac{27648}{15625}.
\end{align}

The reduced E2 matrix elements are
\begin{align}
\langle 3d_{3/2}\|r^2C^{(2)}\|1s_{1/2}\rangle
&= \frac{81\sqrt6}{128},
\\
\langle 3d_{3/2}\|r^2C^{(2)}\|2s_{1/2}\rangle
&= -\frac{10616832\sqrt3}{390625},
\\
\langle 3d_{5/2}\|r^2 C^{(2)}\|1s_{1/2}\rangle
&=
\frac{243}{128},
\\
\langle 3d_{5/2}\|r^2 C^{(2)}\|2s_{1/2}\rangle
&=
-\frac{15925248\sqrt2}{390625}.
\end{align}


\vspace{2ex} 



\begin{thebibliography}{23}%
\makeatletter
\providecommand \@ifxundefined [1]{%
 \@ifx{#1\undefined}
}%
\providecommand \@ifnum [1]{%
 \ifnum #1\expandafter \@firstoftwo
 \else \expandafter \@secondoftwo
 \fi
}%
\providecommand \@ifx [1]{%
 \ifx #1\expandafter \@firstoftwo
 \else \expandafter \@secondoftwo
 \fi
}%
\providecommand \natexlab [1]{#1}%
\providecommand \enquote  [1]{``#1''}%
\providecommand \bibnamefont  [1]{#1}%
\providecommand \bibfnamefont [1]{#1}%
\providecommand \citenamefont [1]{#1}%
\providecommand \href@noop [0]{\@secondoftwo}%
\providecommand \href [0]{\begingroup \@sanitize@url \@href}%
\providecommand \@href[1]{\@@startlink{#1}\@@href}%
\providecommand \@@href[1]{\endgroup#1\@@endlink}%
\providecommand \@sanitize@url [0]{\catcode `\\12\catcode `\$12\catcode
  `\&12\catcode `\#12\catcode `\^12\catcode `\_12\catcode `\%12\relax}%
\providecommand \@@startlink[1]{}%
\providecommand \@@endlink[0]{}%
\providecommand \url  [0]{\begingroup\@sanitize@url \@url }%
\providecommand \@url [1]{\endgroup\@href {#1}{\urlprefix }}%
\providecommand \urlprefix  [0]{URL }%
\providecommand \Eprint [0]{\href }%
\providecommand \doibase [0]{https://doi.org/}%
\providecommand \selectlanguage [0]{\@gobble}%
\providecommand \bibinfo  [0]{\@secondoftwo}%
\providecommand \bibfield  [0]{\@secondoftwo}%
\providecommand \translation [1]{[#1]}%
\providecommand \BibitemOpen [0]{}%
\providecommand \bibitemStop [0]{}%
\providecommand \bibitemNoStop [0]{.\EOS\space}%
\providecommand \EOS [0]{\spacefactor3000\relax}%
\providecommand \BibitemShut  [1]{\csname bibitem#1\endcsname}%
\let\auto@bib@innerbib\@empty
\bibitem [{\citenamefont {Dzuba}\ and\ \citenamefont
  {Flambaum}(2012)}]{DF-PNC12}%
  \BibitemOpen
  \bibfield  {author} {\bibinfo {author} {\bibfnamefont {V.~A.}\ \bibnamefont
  {Dzuba}}\ and\ \bibinfo {author} {\bibfnamefont {V.~V.}\ \bibnamefont
  {Flambaum}},\ }\bibfield  {title} {\bibinfo {title} {Parity violation and
  electric dipole moments in atoms and molecules},\ }\href@noop {} {\bibfield
  {journal} {\bibinfo  {journal} {Int. J. Mod. Phys. E}\ }\textbf {\bibinfo
  {volume} {21}},\ \bibinfo {pages} {1230010} (\bibinfo {year}
  {2012})}\BibitemShut {NoStop}%
\bibitem [{\citenamefont {Roberts}\ \emph {et~al.}(2015)\citenamefont
  {Roberts}, \citenamefont {Dzuba},\ and\ \citenamefont {V.Flambaum}}]{RevPNC}%
  \BibitemOpen
  \bibfield  {author} {\bibinfo {author} {\bibfnamefont {B.~M.}\ \bibnamefont
  {Roberts}}, \bibinfo {author} {\bibfnamefont {V.~A.}\ \bibnamefont {Dzuba}},\
  and\ \bibinfo {author} {\bibfnamefont {V.}~\bibnamefont {V.Flambaum}},\
  }\bibfield  {title} {\bibinfo {title} {Parity and time-reversal violation in
  atomic systems},\ }\href@noop {} {\bibfield  {journal} {\bibinfo  {journal}
  {Annual Rev. Nuc. Part. Science}\ }\textbf {\bibinfo {volume} {65}},\
  \bibinfo {pages} {63} (\bibinfo {year} {2015})}\BibitemShut {NoStop}%
\bibitem [{\citenamefont {Bouchiat}\ and\ \citenamefont
  {Fayet}(2005)}]{BouchiatFayet2005}%
  \BibitemOpen
  \bibfield  {author} {\bibinfo {author} {\bibfnamefont {C.}~\bibnamefont
  {Bouchiat}}\ and\ \bibinfo {author} {\bibfnamefont {P.}~\bibnamefont
  {Fayet}},\ }\bibfield  {title} {\bibinfo {title} {Constraints on the
  parity-violating couplings of a new gauge boson},\ }\href
  {https://doi.org/10.1016/j.physletb.2004.12.065} {\bibfield  {journal}
  {\bibinfo  {journal} {Physics Letters B}\ }\textbf {\bibinfo {volume}
  {608}},\ \bibinfo {pages} {87} (\bibinfo {year} {2005})},\ \Eprint
  {https://arxiv.org/abs/hep-ph/0410260} {arXiv:hep-ph/0410260 [hep-ph]}
  \BibitemShut {NoStop}%
\bibitem [{\citenamefont {Diener}\ \emph {et~al.}(2012)\citenamefont {Diener},
  \citenamefont {Godfrey},\ and\ \citenamefont
  {Turan}}]{DienerGodfreyTuran2012}%
  \BibitemOpen
  \bibfield  {author} {\bibinfo {author} {\bibfnamefont {R.}~\bibnamefont
  {Diener}}, \bibinfo {author} {\bibfnamefont {S.}~\bibnamefont {Godfrey}},\
  and\ \bibinfo {author} {\bibfnamefont {I.}~\bibnamefont {Turan}},\ }\bibfield
   {title} {\bibinfo {title} {Constraining extra neutral gauge bosons with
  atomic parity violation measurements},\ }\href
  {https://doi.org/10.1103/PhysRevD.86.115017} {\bibfield  {journal} {\bibinfo
  {journal} {Physical Review D}\ }\textbf {\bibinfo {volume} {86}},\ \bibinfo
  {pages} {115017} (\bibinfo {year} {2012})},\ \Eprint
  {https://arxiv.org/abs/1111.4566} {arXiv:1111.4566 [hep-ph]} \BibitemShut
  {NoStop}%
\bibitem [{\citenamefont {Dzuba}\ \emph {et~al.}(2017)\citenamefont {Dzuba},
  \citenamefont {Flambaum},\ and\ \citenamefont {Stadnik}}]{DFS17}%
  \BibitemOpen
  \bibfield  {author} {\bibinfo {author} {\bibfnamefont {V.~A.}\ \bibnamefont
  {Dzuba}}, \bibinfo {author} {\bibfnamefont {V.~V.}\ \bibnamefont
  {Flambaum}},\ and\ \bibinfo {author} {\bibfnamefont {Y.~V.}\ \bibnamefont
  {Stadnik}},\ }\bibfield  {title} {\bibinfo {title} {Probing low-mass vector
  bosons with parity nonconservation and nuclear anapole moment measurements in
  atoms and molecules},\ }\href@noop {} {\bibfield  {journal} {\bibinfo
  {journal} {Phys. Rev. Lett.}\ }\textbf {\bibinfo {volume} {119}},\ \bibinfo
  {pages} {223201} (\bibinfo {year} {2017})}\BibitemShut {NoStop}%
\bibitem [{\citenamefont {Nguyen}\ \emph {et~al.}(1997)\citenamefont {Nguyen},
  \citenamefont {Budker}, \citenamefont {DeMille},\ and\ \citenamefont
  {Zolotorev}}]{Dy}%
  \BibitemOpen
  \bibfield  {author} {\bibinfo {author} {\bibfnamefont {A.~T.}\ \bibnamefont
  {Nguyen}}, \bibinfo {author} {\bibfnamefont {D.}~\bibnamefont {Budker}},
  \bibinfo {author} {\bibfnamefont {D.}~\bibnamefont {DeMille}},\ and\ \bibinfo
  {author} {\bibfnamefont {M.}~\bibnamefont {Zolotorev}},\ }\bibfield  {title}
  {\bibinfo {title} {Search for parity nonconservation in atomic dysprosium},\
  }\href {https://doi.org/10.1103/PhysRevA.56.3453} {\bibfield  {journal}
  {\bibinfo  {journal} {Phys. Rev. A}\ }\textbf {\bibinfo {volume} {56}},\
  \bibinfo {pages} {3453} (\bibinfo {year} {1997})}\BibitemShut {NoStop}%
\bibitem [{\citenamefont {Cahn}\ and\ \citenamefont
  {Kane}(1977)}]{CahnKane1977}%
  \BibitemOpen
  \bibfield  {author} {\bibinfo {author} {\bibfnamefont {R.~N.}\ \bibnamefont
  {Cahn}}\ and\ \bibinfo {author} {\bibfnamefont {G.~L.}\ \bibnamefont
  {Kane}},\ }\bibfield  {title} {\bibinfo {title} {Parity violations in
  hydrogen and the fundamental structure of the weak current},\ }\href
  {https://doi.org/10.1016/0370-2693(77)90235-0} {\bibfield  {journal}
  {\bibinfo  {journal} {Physics Letters B}\ }\textbf {\bibinfo {volume} {71}},\
  \bibinfo {pages} {348} (\bibinfo {year} {1977})}\BibitemShut {NoStop}%
\bibitem [{\citenamefont {Dunford}\ and\ \citenamefont
  {Holt}(2007)}]{DunfordHolt2007}%
  \BibitemOpen
  \bibfield  {author} {\bibinfo {author} {\bibfnamefont {R.~W.}\ \bibnamefont
  {Dunford}}\ and\ \bibinfo {author} {\bibfnamefont {R.~J.}\ \bibnamefont
  {Holt}},\ }\bibfield  {title} {\bibinfo {title} {Parity violation in hydrogen
  revisited},\ }\href {https://doi.org/10.1088/0954-3899/34/10/001} {\bibfield
  {journal} {\bibinfo  {journal} {Journal of Physics G: Nuclear and Particle
  Physics}\ }\textbf {\bibinfo {volume} {34}},\ \bibinfo {pages} {2099}
  (\bibinfo {year} {2007})},\ \Eprint {https://arxiv.org/abs/0706.2407}
  {arXiv:0706.2407 [hep-ph]} \BibitemShut {NoStop}%
\bibitem [{\citenamefont {Dunford}\ and\ \citenamefont
  {Holt}(2011)}]{DunfordHolt2011}%
  \BibitemOpen
  \bibfield  {author} {\bibinfo {author} {\bibfnamefont {R.~W.}\ \bibnamefont
  {Dunford}}\ and\ \bibinfo {author} {\bibfnamefont {R.~J.}\ \bibnamefont
  {Holt}},\ }\bibfield  {title} {\bibinfo {title} {Parity nonconservation in
  hydrogen},\ }\href {https://doi.org/10.1007/s10751-011-0278-8} {\bibfield
  {journal} {\bibinfo  {journal} {Hyperfine Interactions}\ }\textbf {\bibinfo
  {volume} {200}},\ \bibinfo {pages} {45} (\bibinfo {year} {2011})}\BibitemShut
  {NoStop}%
\bibitem [{\citenamefont {Rasor}\ and\ \citenamefont
  {Yost}(2020)}]{RasorYost2020}%
  \BibitemOpen
  \bibfield  {author} {\bibinfo {author} {\bibfnamefont {C.}~\bibnamefont
  {Rasor}}\ and\ \bibinfo {author} {\bibfnamefont {D.~C.}\ \bibnamefont
  {Yost}},\ }\bibfield  {title} {\bibinfo {title} {Laser-based measurement of
  parity violation in hydrogen},\ }\href
  {https://doi.org/10.1103/PhysRevA.102.032801} {\bibfield  {journal} {\bibinfo
   {journal} {Physical Review A}\ }\textbf {\bibinfo {volume} {102}},\ \bibinfo
  {pages} {032801} (\bibinfo {year} {2020})}\BibitemShut {NoStop}%
\bibitem [{\citenamefont {Li}\ \emph {et~al.}(2024)\citenamefont {Li},
  \citenamefont {Derevianko},\ and\ \citenamefont
  {Elliott}}]{LiDereviankoElliott2024}%
  \BibitemOpen
  \bibfield  {author} {\bibinfo {author} {\bibfnamefont {J.}~\bibnamefont
  {Li}}, \bibinfo {author} {\bibfnamefont {A.}~\bibnamefont {Derevianko}},\
  and\ \bibinfo {author} {\bibfnamefont {D.~S.}\ \bibnamefont {Elliott}},\
  }\bibfield  {title} {\bibinfo {title} {Feasibility of extracting the proton
  weak charge from quantum-control measurements of atomic parity violation on
  the {$2s$--$3s$} or {$2s$--$4s$} transition in hydrogen},\ }\href
  {https://doi.org/10.1103/PhysRevA.109.012808} {\bibfield  {journal} {\bibinfo
   {journal} {Physical Review A}\ }\textbf {\bibinfo {volume} {109}},\ \bibinfo
  {pages} {012808} (\bibinfo {year} {2024})},\ \Eprint
  {https://arxiv.org/abs/2310.14689} {arXiv:2310.14689 [physics.atom-ph]}
  \BibitemShut {NoStop}%
\bibitem [{\citenamefont {Erler}\ \emph {et~al.}(2003)\citenamefont {Erler},
  \citenamefont {Kurylov},\ and\ \citenamefont
  {Ramsey-Musolf}}]{ErlerKurylovRamseyMusolf2003}%
  \BibitemOpen
  \bibfield  {author} {\bibinfo {author} {\bibfnamefont {J.}~\bibnamefont
  {Erler}}, \bibinfo {author} {\bibfnamefont {A.}~\bibnamefont {Kurylov}},\
  and\ \bibinfo {author} {\bibfnamefont {M.~J.}\ \bibnamefont
  {Ramsey-Musolf}},\ }\bibfield  {title} {\bibinfo {title} {Weak charge of the
  proton and new physics},\ }\href {https://doi.org/10.1103/PhysRevD.68.016006}
  {\bibfield  {journal} {\bibinfo  {journal} {Physical Review D}\ }\textbf
  {\bibinfo {volume} {68}},\ \bibinfo {pages} {016006} (\bibinfo {year}
  {2003})},\ \Eprint {https://arxiv.org/abs/hep-ph/0302149}
  {arXiv:hep-ph/0302149 [hep-ph]} \BibitemShut {NoStop}%
\bibitem [{\citenamefont {Androi{\'c}}\ \emph {et~al.}(2018)\citenamefont
  {Androi{\'c}} \emph {et~al.}}]{AndroicEtAl2018}%
  \BibitemOpen
  \bibfield  {author} {\bibinfo {author} {\bibfnamefont {D.}~\bibnamefont
  {Androi{\'c}}} \emph {et~al.} (\bibinfo {collaboration} {Jefferson Lab Qweak
  Collaboration}),\ }\bibfield  {title} {\bibinfo {title} {Precision
  measurement of the weak charge of the proton},\ }\href
  {https://doi.org/10.1038/s41586-018-0096-0} {\bibfield  {journal} {\bibinfo
  {journal} {Nature}\ }\textbf {\bibinfo {volume} {557}},\ \bibinfo {pages}
  {207} (\bibinfo {year} {2018})},\ \Eprint {https://arxiv.org/abs/1905.08283}
  {arXiv:1905.08283 [nucl-ex]} \BibitemShut {NoStop}%
\bibitem [{\citenamefont {Carlini}\ \emph {et~al.}(2019)\citenamefont
  {Carlini}, \citenamefont {van Oers}, \citenamefont {Pitt},\ and\
  \citenamefont {Smith}}]{CarliniVanOersPittSmith2019}%
  \BibitemOpen
  \bibfield  {author} {\bibinfo {author} {\bibfnamefont {R.~D.}\ \bibnamefont
  {Carlini}}, \bibinfo {author} {\bibfnamefont {W.~T.~H.}\ \bibnamefont {van
  Oers}}, \bibinfo {author} {\bibfnamefont {M.~L.}\ \bibnamefont {Pitt}},\ and\
  \bibinfo {author} {\bibfnamefont {G.~R.}\ \bibnamefont {Smith}},\ }\bibfield
  {title} {\bibinfo {title} {Determination of the proton's weak charge and its
  constraints on the standard model},\ }\href
  {https://doi.org/10.1146/annurev-nucl-101918-023633} {\bibfield  {journal}
  {\bibinfo  {journal} {Annual Review of Nuclear and Particle Science}\
  }\textbf {\bibinfo {volume} {69}},\ \bibinfo {pages} {191} (\bibinfo {year}
  {2019})}\BibitemShut {NoStop}%
\bibitem [{\citenamefont {Dev}\ \emph {et~al.}(2021)\citenamefont {Dev},
  \citenamefont {Rodejohann}, \citenamefont {Xu},\ and\ \citenamefont
  {Zhang}}]{DevRodejohannXuZhang2021}%
  \BibitemOpen
  \bibfield  {author} {\bibinfo {author} {\bibfnamefont {P.~S.~B.}\
  \bibnamefont {Dev}}, \bibinfo {author} {\bibfnamefont {W.}~\bibnamefont
  {Rodejohann}}, \bibinfo {author} {\bibfnamefont {X.-J.}\ \bibnamefont {Xu}},\
  and\ \bibinfo {author} {\bibfnamefont {Y.}~\bibnamefont {Zhang}},\ }\bibfield
   {title} {\bibinfo {title} {Searching for {$Z'$} bosons at the {P2}
  experiment},\ }\href {https://doi.org/10.1007/JHEP06(2021)039} {\bibfield
  {journal} {\bibinfo  {journal} {Journal of High Energy Physics}\ }\textbf
  {\bibinfo {volume} {2021}},\ \bibinfo {pages} {039} (\bibinfo {year}
  {2021})},\ \Eprint {https://arxiv.org/abs/2103.09067} {arXiv:2103.09067
  [hep-ph]} \BibitemShut {NoStop}%
\bibitem [{\citenamefont {Thomas}\ \emph {et~al.}(2022)\citenamefont {Thomas},
  \citenamefont {Wang},\ and\ \citenamefont
  {Williams}}]{ThomasWangWilliams2022}%
  \BibitemOpen
  \bibfield  {author} {\bibinfo {author} {\bibfnamefont {A.~W.}\ \bibnamefont
  {Thomas}}, \bibinfo {author} {\bibfnamefont {X.~G.}\ \bibnamefont {Wang}},\
  and\ \bibinfo {author} {\bibfnamefont {A.~G.}\ \bibnamefont {Williams}},\
  }\bibfield  {title} {\bibinfo {title} {Sensitivity of parity-violating
  electron scattering to a dark photon},\ }\href
  {https://doi.org/10.1103/PhysRevLett.129.011807} {\bibfield  {journal}
  {\bibinfo  {journal} {Physical Review Letters}\ }\textbf {\bibinfo {volume}
  {129}},\ \bibinfo {pages} {011807} (\bibinfo {year} {2022})},\ \Eprint
  {https://arxiv.org/abs/2201.06760} {arXiv:2201.06760 [hep-ph]} \BibitemShut
  {NoStop}%
\bibitem [{\citenamefont {Thomas}\ \emph {et~al.}(2025)\citenamefont {Thomas},
  \citenamefont {Wang},\ and\ \citenamefont
  {Williams}}]{ThomasWangWilliams2025}%
  \BibitemOpen
  \bibfield  {author} {\bibinfo {author} {\bibfnamefont {A.~W.}\ \bibnamefont
  {Thomas}}, \bibinfo {author} {\bibfnamefont {X.~G.}\ \bibnamefont {Wang}},\
  and\ \bibinfo {author} {\bibfnamefont {A.~G.}\ \bibnamefont {Williams}},\
  }\bibfield  {title} {\bibinfo {title} {Dark photon in parity-violating
  electron scatterings},\ }\href@noop {} {\bibfield  {journal} {\bibinfo
  {journal} {arXiv e-prints}\ } (\bibinfo {year} {2025})},\ \bibinfo {note}
  {preprint},\ \Eprint {https://arxiv.org/abs/2505.07279} {arXiv:2505.07279
  [hep-ph]} \BibitemShut {NoStop}%
\bibitem [{\citenamefont {Dzuba}\ \emph
  {et~al.}(2026{\natexlab{a}})\citenamefont {Dzuba}, \citenamefont {Flambaum},\
  and\ \citenamefont {Vong}}]{Sr}%
  \BibitemOpen
  \bibfield  {author} {\bibinfo {author} {\bibfnamefont {V.~A.}\ \bibnamefont
  {Dzuba}}, \bibinfo {author} {\bibfnamefont {V.~V.}\ \bibnamefont
  {Flambaum}},\ and\ \bibinfo {author} {\bibfnamefont {G.~K.}\ \bibnamefont
  {Vong}},\ }\bibfield  {title} {\bibinfo {title} {Parity nonconservation in rb
  and ${\mathrm{sr}}^{+}$ due to a low-mass vector boson},\ }\href
  {https://doi.org/10.1103/qf7h-drc2} {\bibfield  {journal} {\bibinfo
  {journal} {Phys. Rev. A}\ }\textbf {\bibinfo {volume} {113}},\ \bibinfo
  {pages} {052808} (\bibinfo {year} {2026}{\natexlab{a}})}\BibitemShut
  {NoStop}%
\bibitem [{\citenamefont {Dzuba}\ \emph
  {et~al.}(2026{\natexlab{b}})\citenamefont {Dzuba}, \citenamefont {Flambaum},\
  and\ \citenamefont {Vong}}]{HPV}%
  \BibitemOpen
  \bibfield  {author} {\bibinfo {author} {\bibfnamefont {V.~A.}\ \bibnamefont
  {Dzuba}}, \bibinfo {author} {\bibfnamefont {V.~V.}\ \bibnamefont
  {Flambaum}},\ and\ \bibinfo {author} {\bibfnamefont {G.~K.}\ \bibnamefont
  {Vong}},\ }\bibfield  {title} {\bibinfo {title} {Parity nonconservation in
  hydrogen induced by low-mass vector-boson exchange},\ }\href
  {https://doi.org/10.1103/pcyj-nkf2} {\bibfield  {journal} {\bibinfo
  {journal} {Phys. Rev. A}\ }\textbf {\bibinfo {volume} {113}},\ \bibinfo
  {pages} {062817} (\bibinfo {year} {2026}{\natexlab{b}})},\ \Eprint
  {https://arxiv.org/abs/26051.1032} {arXiv:26051.1032} \BibitemShut {NoStop}%
\bibitem [{\citenamefont {Tanabashi}\ \emph {et~al.}(2018)\citenamefont
  {Tanabashi}, \citenamefont {Hagiwara}, \citenamefont {Hikasa},\ and\
  \citenamefont {{\em et al}}}]{SM}%
  \BibitemOpen
  \bibfield  {author} {\bibinfo {author} {\bibfnamefont {M.}~\bibnamefont
  {Tanabashi}}, \bibinfo {author} {\bibfnamefont {K.}~\bibnamefont {Hagiwara}},
  \bibinfo {author} {\bibfnamefont {K.}~\bibnamefont {Hikasa}},\ and\ \bibinfo
  {author} {\bibnamefont {{\em et al}}} (\bibinfo {collaboration} {Particle
  Data Group}),\ }\bibfield  {title} {\bibinfo {title} {Review of particle
  physics},\ }\href {https://doi.org/10.1103/PhysRevD.98.030001} {\bibfield
  {journal} {\bibinfo  {journal} {Phys. Rev. D}\ }\textbf {\bibinfo {volume}
  {98}},\ \bibinfo {pages} {030001} (\bibinfo {year} {2018})}\BibitemShut
  {NoStop}%
\bibitem [{\citenamefont {Flambaum}\ and\ \citenamefont
  {Samsonov}(2026)}]{Samsonov}%
  \BibitemOpen
  \bibfield  {author} {\bibinfo {author} {\bibfnamefont {V.~V.}\ \bibnamefont
  {Flambaum}}\ and\ \bibinfo {author} {\bibfnamefont {I.~B.}\ \bibnamefont
  {Samsonov}},\ }\bibfield  {title} {\bibinfo {title} {Effects of dispersion
  parity-violating interaction in electron scattering and atoms},\ }\href
  {https://doi.org/10.1103/cb8l-bt9n} {\bibfield  {journal} {\bibinfo
  {journal} {Phys. Rev. D}\ }\textbf {\bibinfo {volume} {114}},\ \bibinfo
  {pages} {L011302} (\bibinfo {year} {2026})},\ \Eprint
  {https://arxiv.org/abs/2602.22466} {arXiv:2602.22466 [hep-ph]} \BibitemShut
  {NoStop}%
\bibitem [{\citenamefont {Flambaum}\ and\ \citenamefont
  {Khriplovich}(1980)}]{anapole}%
  \BibitemOpen
  \bibfield  {author} {\bibinfo {author} {\bibfnamefont {V.~V.}\ \bibnamefont
  {Flambaum}}\ and\ \bibinfo {author} {\bibfnamefont {I.~B.}\ \bibnamefont
  {Khriplovich}},\ }\bibfield  {title} {\bibinfo {title} {P-odd nuclear forces
  - a source of parity violation in atoms},\ }\href@noop {} {\bibfield
  {journal} {\bibinfo  {journal} {Sov. Phys. JETP}\ }\textbf {\bibinfo {volume}
  {52}},\ \bibinfo {pages} {835} (\bibinfo {year} {1980})}\BibitemShut
  {NoStop}%
\bibitem [{\citenamefont {Bethe}\ and\ \citenamefont
  {Salpeter}(1957)}]{BetheSalpeter1957}%
  \BibitemOpen
  \bibfield  {author} {\bibinfo {author} {\bibfnamefont {H.~A.}\ \bibnamefont
  {Bethe}}\ and\ \bibinfo {author} {\bibfnamefont {E.~E.}\ \bibnamefont
  {Salpeter}},\ }\href@noop {} {\emph {\bibinfo {title} {Quantum Mechanics of
  One- and Two-Electron Atoms}}}\ (\bibinfo  {publisher} {Springer-Verlag},\
  \bibinfo {address} {Berlin},\ \bibinfo {year} {1957})\BibitemShut {NoStop}%
\end{thebibliography}

%


\end{document}